**Simulation of Noncircular Rigid Bodies: Machine Learning Based Overlap Calculation Technique with System Size Independent Computational Cost**

*Saientan Bag**

Eduard-Zintl-Institut für Anorganische und Physikalische Chemie, Technische Universität Darmstadt, Alarich-Weiss-Str. 8, 64287 Darmstadt, Germany.

Institute of Nanotechnology (INT), Karlsruhe Institute of Technology (KIT), Eggenstein Leopoldshafen 76344, Germany.

*email: bag.physics@gmail.com

## Abstract

Standard molecular dynamics (MD) and Monte Carlo (MC) simulation deals with spherical particles. Extending these standard simulation methodologies to the non-spherical cases is non-trivial. To circumvent this problem, non-spherical bodies are considered as a collection of constituent spherical objects. As the number of these constituent objects becomes large, the computational burden to simulate the system also increases. In this article, we propose an alternative way to simulate non-circular rigid bodies in two dimensions having pairwise repulsive interactions. Our approach is based on a machine learning (ML) based model which predicts the overlap between two non-circular bodies. The machine learning model is easy to train and the computation cost of its implementation remains independent of the number of constituents disks used to represent a non-circular rigid body. When used in MC simulation, our approach provides significant speed up in comparison to the standard implementation where overlap determination between two rigid bodies is done by calculating the distance of their constituent disks. Our proposed ML based MC method provided very similar structural features (in comparison to standard implementation) of the systems. We believe this work is a very first step towards a time-efficient simulation of non-spherical rigid bodies.

## Introduction

In the last 50 years, we have seen the enormous development of simulation techniques in terms of Molecular Dynamics (MD) and Monte Carlo (MC) simulation[1]. The success of these extremely powerful simulation methodologies to calculate the structural and dynamical properties of various systems is unparalleled[2]. However, the standard MD and MC simulations mostly deal with the spherical particles. While the simulation of non-spherical particles is relevant in a great variety of contexts e.g. to study the effect of crowding in various biological processes[3], to understand the self-assembly of non-spherical magnetic particle[4] for biotechnological applications. Extending the MD and MC simulation for non-spherical particles is not straightforward. Even for a very simple case of pairwise hard repulsion, there is no simple way to simulate non spherical particles because of the lack of overlap detection methodologies between them. Except for a few standard cases, there is no exact way to determine overlap between two rigid bodies. For example, in the case of a rigid spherocylinder, an analytical formula[5] for the overlap determination can be written by considering the minimum distance between two line segments. This problem (overlap determination for non-spherical bodies) is circumvented by considering the rigid body as a collection of spherical particles[6] (or disks in two dimensions) and calculating the overlap between the spherical particle (or disk-like particles). However, this method of overlap calculation becomes very expensive as the number of constituents spheres (or disks) increases which makes the simulation unfeasible. In this paper, we propose a data driven approach to overcome this problem. We train a Machine Learning (ML) model to determine whether there is overlap between two rigid bodies given the position and orientation of the rigid bodies. Therefore, the overlap detection time remains independent of the number of constituent spheres (or disk). This allows simulations of large collections of rigid bodies for a longer time. In this paper, we considered rigid bodies of four different shapes (Concentric Circle, Triangle, Rod and Star) and explained the ML model building in great detail. We further perform MC simulation with this ML model (to detect the overlap) and compare the results with the actual simulation. We believe that this will generate interest in the molecular simulation community as an alternative way to simulate collection of non-spherical rigid bodies.

## Methodology and Results

We considered rigid bodies of four different shapes namely concentric circle, triangle, rod and star (see Figure 1). To calculate whether there is an overlap between two bodies, the rigid bodies are modeled as many constituent small disks of the same size as shown in Figure 1 below. To represent bodies of different shapes different numbers of disks are required. As the size of the disk becomes smaller, the more accurately they would mimic the peripheral shape of the rigid body (see Figure 1).

33

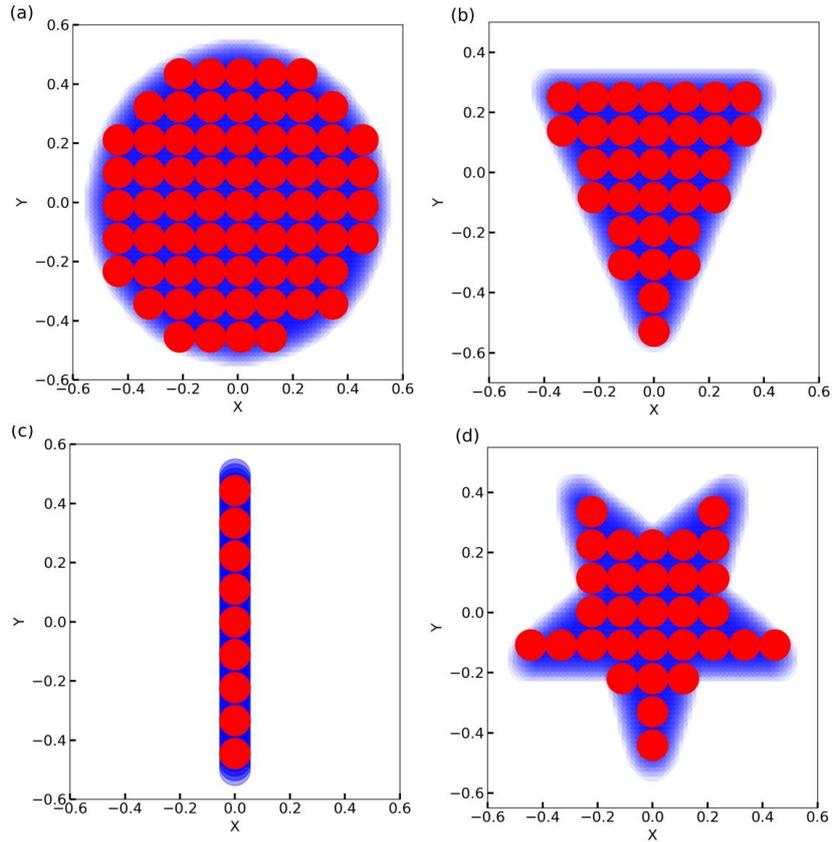

Figure 1: (a-d) Rigid body (the blue background) with different geometric shapes modelled by a series of small disks (shown in red). Number of disks required to represent the (a) concentric circle, (b) triangle (c) rod and (d) Star are 67, 32, 9 and 31 respectively.

The diameter of the concentric circle was 1 unit of length. The long axis of all the other three rigid bodies were also roughly 1 unit in length. To calculate whether there is an overlap between two bodies A and B, we calculate the distance between all disk pairs with one disk belonging to body A and another belonging to body B. If there are $N$ constituent disks for a rigid body, then these requires $N^2$ distance calculation. Now if any of these $N^2$ distances is lower than a certain $cutoff$, then we say that the body A and B are overlapped. Since this overlap estimation requires $N^2$ distance calculations, for large $N$, the computation time grows rapidly. Therefore, we designed a ML model to predict the overlap between two bodies. The input and output of the ML model are described in Figure 2 below.

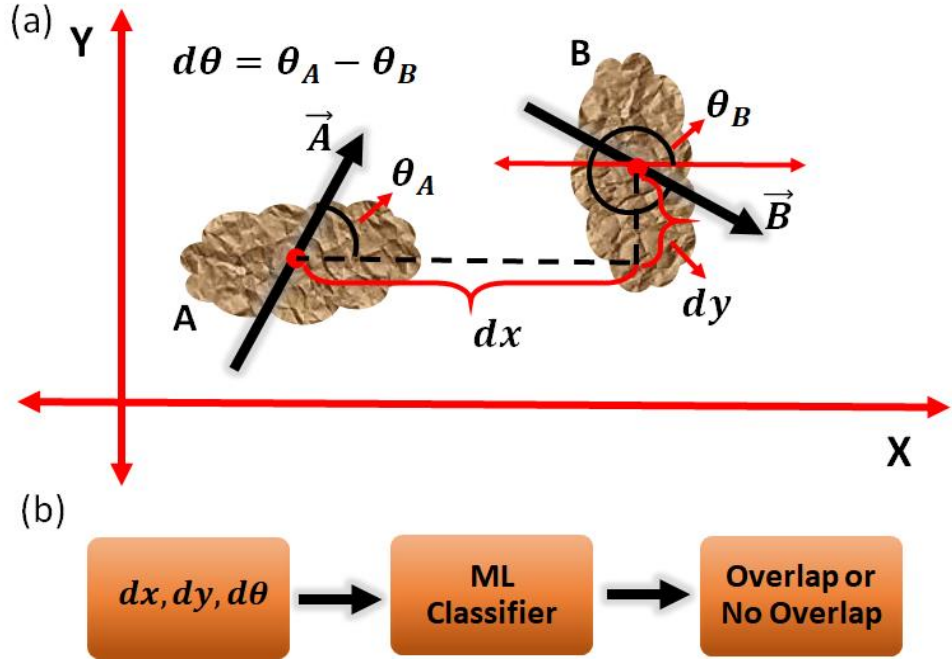

Figure 2: (a-b) Schematic diagram describing the input and output of a ML classifier model to predict the overlap between two rigid bodies A and B in two-dimension. (a) The position and the orientation of the body is represented by the center of geometry (shown as red dot) and a vector $\vec{A}$ attached with the rigid body in any arbitrary direction. As the body moves and rotates, the vector $\vec{A}$ moves and rotates with it. (b) The relative distance between the center of geometries and the relative angle between these body fixed vectors ($\vec{A}$ and $\vec{B}$) are used to determine whether there is an overlap between them.

To generate the dataset for the ML model we randomly kept two rigid bodies at different random relative position and orientation and determined the overlap by calculating the distance between the constituent disks representing the rigid body (see Figure 2). The $cutoff$ distance for the overlap determination was 0.11 which is also the distance between two neighboring constituent disk (see red disks in Figure 1.) in a body. The relative distance ($dx, dy$) was kept between 0 to 10 in both x and y direction. The relative orientation $d\theta$ (see Figure 2) is between 0 to $2\pi$. We train ML classification models (see Figure 2) which takes $dx, dy$ and $d\theta$ as input and predict whether there is overlap or not. Therefore, this is a binary classification problem.

To generate the learning curve for the ML models we keep on increasing the training data set and measure the prediction accuracy of 2000 test data points. In both the training and test data set we kept an equal number of data of two categories (overlap and no overlap). We used the "Gradient Boosting Classifier" as a ML model here. However, there is nothing special about this ML model. Any other accurate and fast classification model should do the job. To implement the "Gradient Boosting Classification" models, we used the "Sklearn"[7] software module with all the default settings. The learning curves for the 4 cases are shown in Figure 3 below.

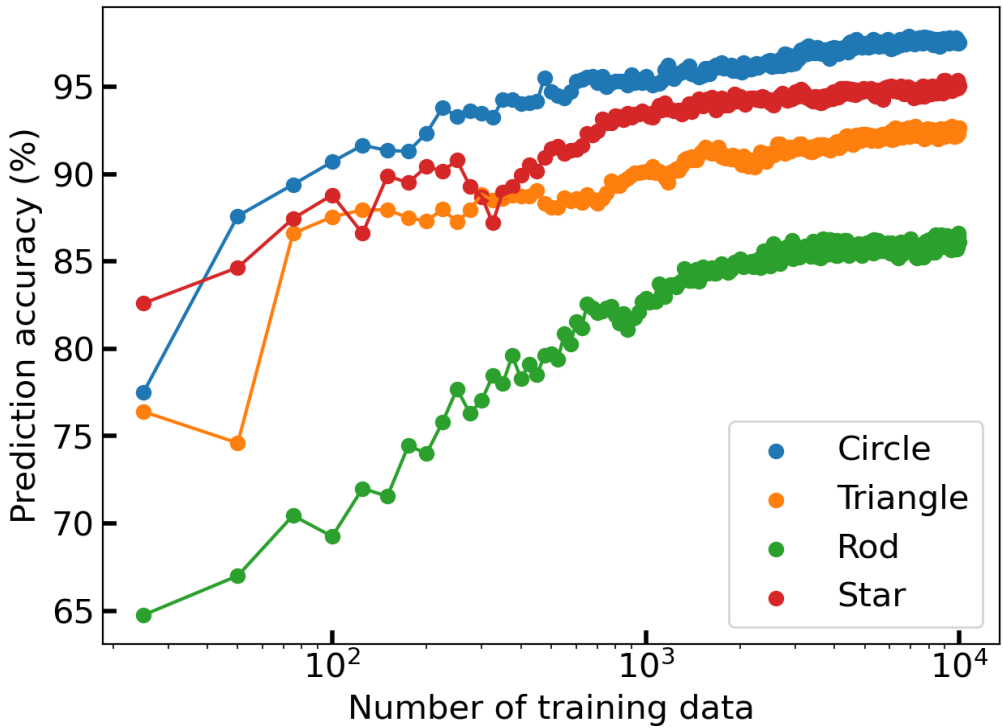

Figure 3: Overlap prediction accuracy in test dataset as a function of number of training data for different rigid body cases. The accuracy is the highest for the circle case (98 %) and the lowest (86 %) for the rod. The cases for star and triangle lie in between with prediction accuracy of 95 and 92 respectively.

It is evident from Figure 3 that the performance of the ML model is best for the "Circle" case with 98 % accuracy in predicting the test cases. This high accuracy is expected because the ML model can simply learn the distance between the centers of geometry to determine whether there is overlap or not. However, as the rigid body becomes more circularly asymmetric, the ML model becomes more inaccurate with the lowest accuracy of 86% for the "rod" case. The case for "star" and "triangle" lie in between with prediction accuracy of 95 and 92 respectively. The "Star" which is more circularly symmetric than "Triangle" generates a better ML model compared to "Triangle". To understand the ML predictions more deeply we calculate the confusion matrix for the test predictions with 2000 test data points (1000 in each category) as below (see Figure 4). In case of a concentric circle, the ML model almost perfectly categorizes the "Overlap" and "No Overlap" with slightly better performance in predicting "No Overlap" in comparison to "Overlap". In all the other 3 cases, the ML model makes more mistakes identifying "No Overlap" to be "Overlap" rather than "Overlap" to be "No Overlap".

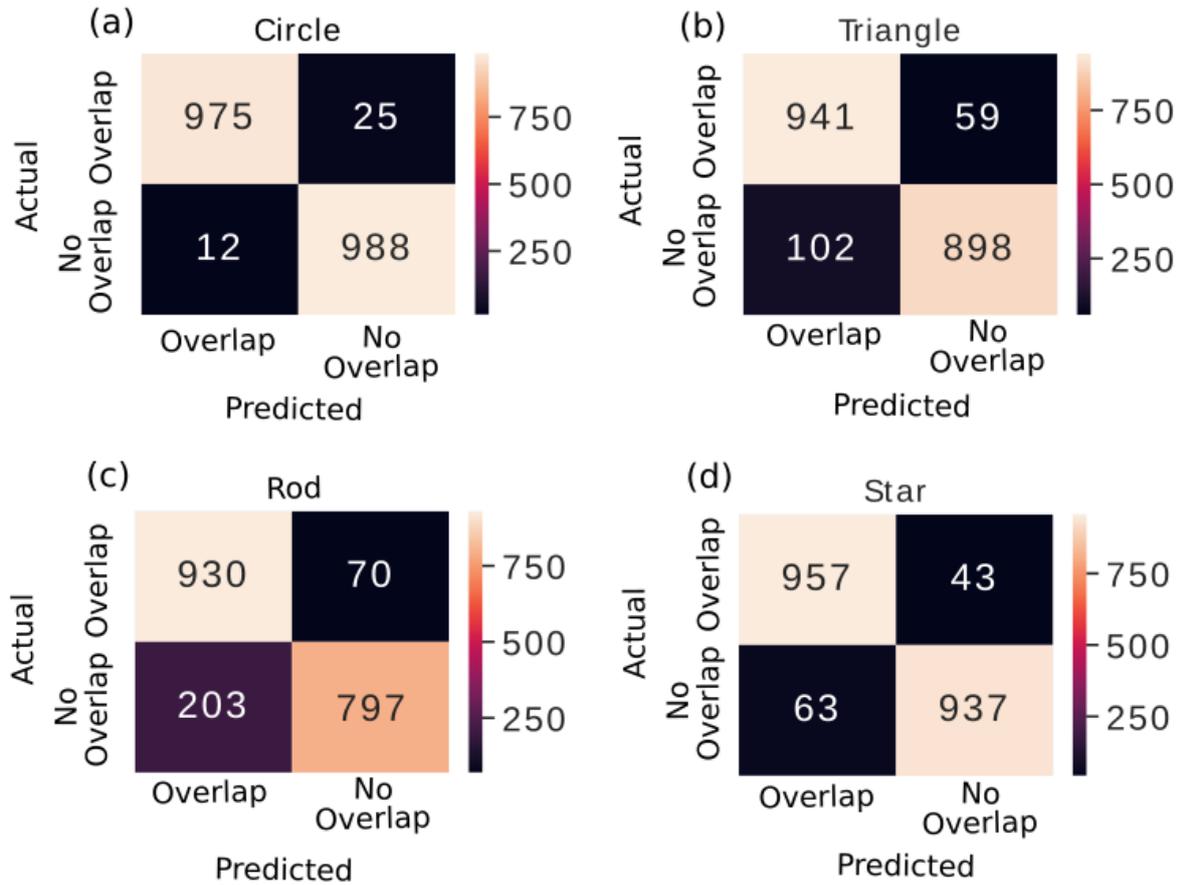

Figure 4: (a-d) Confusion matrices in prediction of the 2000 test data points having 1000 points in each category: "Overlap" and "No Overlap" . (a) In case of a concentric circle, the ML model almost perfectly categorizes the "Overlap" and "No Overlap" with slightly better performance in predicting "No Overlap" in comparison to "Overlap". (b-d) In all other 3 cases, "Overlap" is predicted well than "No overlap" which is more often wrongly predicted as "Overlap".

So far we have shown our approach in designing the ML model and its prediction accuracy. However, for the ML model to be usable at all it needs to provide sufficient advantages in terms of "computation time", otherwise it will be useless. To check that we generate rigid bodies with varying sizes (the shape was fixed to be "rod" for simplicity) and calculate the overlap between them. We calculated the overlap by explicitly calculating the distance of the constituent disks as well as by training a ML model as described in the previous section. The comparison of computation time as a function of size of the body is shown in Figure 5 below. The computation time for the ML methods remains constant while in the other case it almost increases linearly. Although this comparison roughly gives an idea how the ML model can be computationally beneficial to use, it still does not provide what will be its implication in actual simulations where one need to perform many overlap calculations between many pairs.

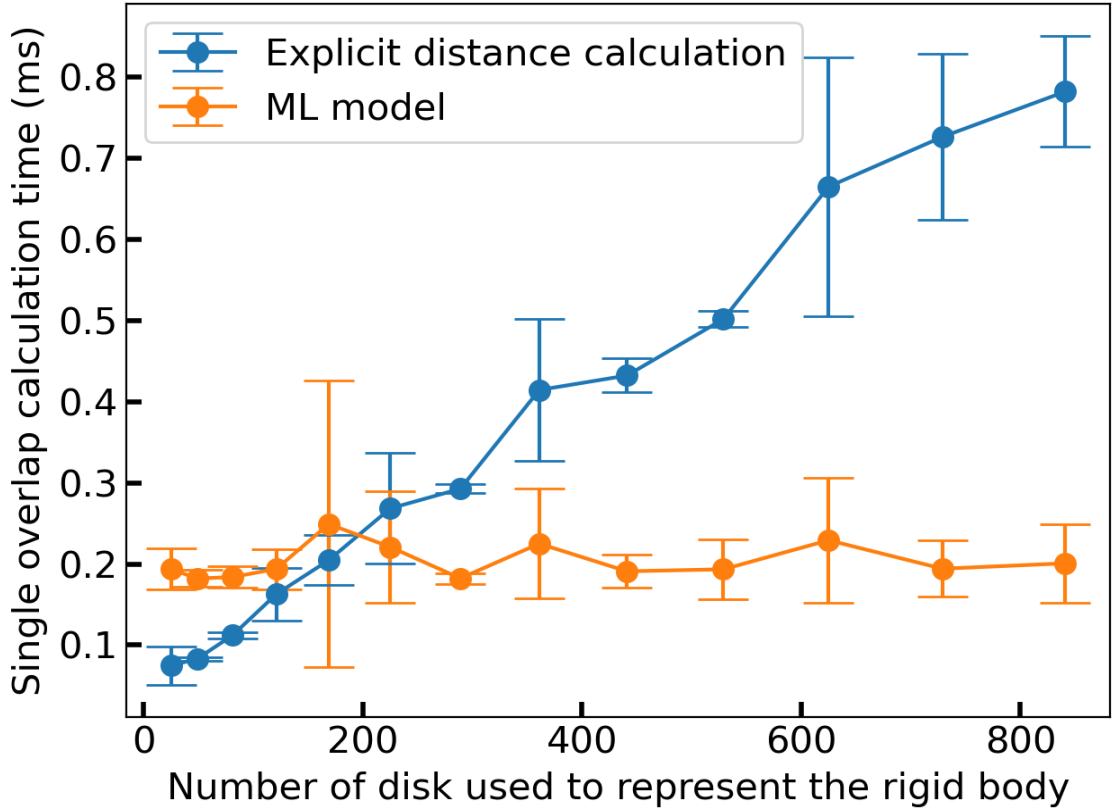

Figure 5: Comparison of single overlap calculation time (in a single cpu) between two rigid bodies as a function of the number of disks used to represent them. The calculation time almost linearly increases with the "number of disks" when estimated using "explicit distance calculation" while the time for the ML model prediction remains constant.

To estimate the implication of the ML method in actual simulation in terms of computational time as well as in predicting the structural properties we performed Monte Carlo (MC) simulations with the following details: We took a system of 64 rigid bodies of the same type and randomly arranged them in a square simulation box. The system was prepared at a number density of 0.55, which resulted in a simulation box length of 10.7. The simulation box was periodic in both x and y direction. During the random arrangement, we made sure that the rigid bodies do not overlap. With this initial random arrangement, we randomly pick a rigid body and propose a MC move. A MC move is a combination of translation (of amount $tx$ and $ty$ in x and y direction) and a rotation (of amount $r\theta$) of the rigid body. Here, $tx$ and $ty$ are chosen to be a random number between -0.4 to +0.4 while $r\theta$ is randomly chosen between -10 to 10 degrees. If this new position and orientation of the body causes overlaps with the other rigid bodies then we discard this move keeping the body to its previous position and orientation. In case of "No overlap", this move is accepted. We repeat this and perform $10^5$ MC steps. Two different set MC simulations are done by using the ML model to determine the overlap and by explicitly calculating the distance between the constituent disks to determine overlap. We repeated the above MC simulations for all the four different types of rigid bodies and compared the simulation time in Table 1 below.

| Rigid Body Type | Total number of disks required to represent the 64 rigid body | Time taken (single cpu) for $10^5$ MC steps : explicit distance calculation | Time taken (single cpu) for $10^5$ MC steps : ML model |
|---|---|---|---|
| Concentric Circle | 4288 | 24481 s | 916 s |
| Triangle | 2048 | 5913 s | 951 s |
| Rod | 576 | 655 s | 956 s |
| Star | 1984 | 5424 s | 1006 s |

Table 1: Comparison of cpu time taken to perform $10^5$ MC steps for 64 rigid bodies.

As expected the simulation time for the ML models almost remains unchanged for 4 cases while in case of "Explicit Distance Calculation" it largely varies as the total number of disks (used to represent the rigid body) changes. In the case of "Circle", we have an ML model which provides 30 times faster MC simulation. In the case of "Star" and "Triangle", the ML model guided MC is roughly 6 times faster. In the rod case the computation time is more or less the same since only 576 disks are required to represent 64 rigid rods. Now, it is quite clear that there is a benefit in terms of computational time in using the ML based overlap determination. Therefore, we now check how comparable are the MC generated structural properties of the systems in these two cases: ML based overlap determination vs explicit distance calculation. The comparison is presented in Figure 6 below. From the generated MC trajectory we calculated the pair correlation function $g(r)$ defined as

$$g(r) = \frac{1}{2\pi r} < \sum_i^N \sum_j^N (\delta(r - r_{ij})) >$$

Here, $r_{ij}$ is the distance between the centers of geometry of two rigid bodies and $N$ is the total number of rigid bodies.

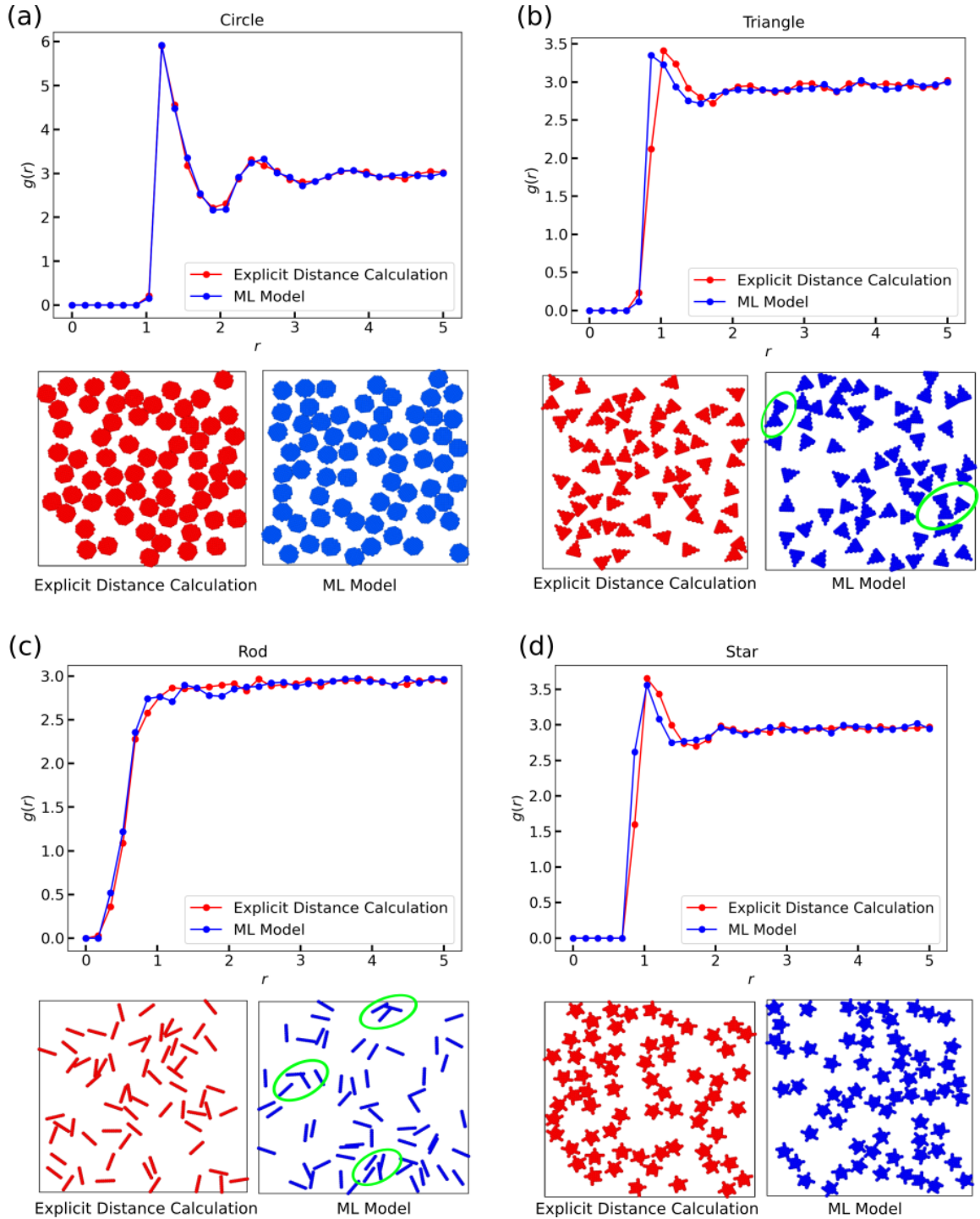

Figure 6: (a-d) Pair correlation function (top panel) and the equilibrated simulation snapshots (bottom panels) obtained from $10^5$ steps of MC Simulation with 64 rigid bodies. MC simulations were performed using the ML model for overlap determination as well as the using explicit distance calculation. (a) The ML model being quite accurate for the concentric circle case generates almost identical $g(r)$ and the snapshots as the "explicit distance calculation" cases. (b) In the triangle particle case, the ML model allows the particle to overlap a little which is reflected in the simulation snapshot (highlighted in green) as well as in $g(r)$ moved towards lower $r$ for the ML case. (c) The rods frequently overlap (highlighted in green) as seen in the snapshots because of the imperfect ML model. Although $g(r)$ is almost unaffected. (d) The snapshots and $g(r)$ are very similar in cases of star shaped particles.

We also show the simulation snapshots (see bottom panels of Figure 6) after the $10^5$ MC steps in both the cases. The ML model is quite accurate for the concentric circle case and therefore generate almost identical $g(r)$ and the snapshots as the "explicit distance calculation" case. In the triangle particle case (see Figure 6(b)), the ML model allows the particle to overlap a little which is reflected in the simulation snapshot (see Figure 6(b) bottom panel) as well as in $g(r)$ which moved towards lower $r$ for the ML case. As shown in the snapshot of Figure 6(c), the rods frequently overlap (highlighted in green in the snapshots) as seen in the snapshots because of the imperfect ML model. Although $g(r)$ is almost unaffected. Figure 6(d) shows that the snapshots and $g(r)$ are very similar (ML vs explicit distance calculation) in cases of star shaped particles.

We have just measured the pair correlation functions here at a particular density and admit that we could possibly measure the other structural properties (at different densities) and there might be larger differences between two models (ML vs explicit distance calculation). For example, in the case of "rod" shaped bodies one could measure the orientational order parameter of the system and check the influence of the ML models.

## Conclusion

MD and MC simulation[1,8] are extremely powerful tools to study structural and dynamical properties of a great variety of physical systems. Enormous amounts of effort have been made in the last few years to make these simulation methodologies more accurate and computationally cheaper. However, the traditional simulation methodologies mostly deal with spherical particles and therefore misses out a great variety of physical systems involving non spherical particles. To bring the non-spherical particle in the same umbrella of established simulation techniques, the non-spherical objects are represented as a collection of constituent spherical beads. But this increases the total number of particles in the system and the corresponding computational costs. In this paper we showed an alternative ML based approach to circumvent this problem. Considering the simple case of excluded volume interaction we performed MC simulation (in two dimension) and compared the resultant structure obtained from ML based approach to the exact ones. The method provides significant speed up for most of the cases (except the rod like object) considered. The resulting structures obtained are also very similar for both the cases. In future, we will generalize this approach to include other pairwise interactions going beyond the excluded volume interaction. Improved performance (both in terms of accuracy and speed) of the ML models will be subject of our future research.

# Acknowledgements

I acknowledge financial support from TU Darmstadt in terms of "Career Bridging Grant" funded by the Hessian Ministry of Science and Arts (HMWK).